\documentclass{article}
\usepackage{spconf,amsmath,graphicx,multirow,booktabs,amssymb,bbding,pifont}
\usepackage[noend]{algpseudocode}
\usepackage[numbers,sort&compress]{natbib}

\title{AUDIO-VISUAL MULTI-CHANNEL SPEECH SEPARATION, DEREVERBERATION AND RECOGNITION}
\name{Guinan Li$^{*,1}$, Jianwei Yu$^{*,1,2}$, Jiajun Deng$^{1}$, Xunying Liu$^{1}$, Helen Meng$^{1}$\thanks{ $^{*}$ Equal contribution; This work is partly done when Jianwei Yu is an intern in Tencent AI lab.}}
\address{$^1$The Chinese University of Hong Kong; $^2$Tencent AI lab }
\newcommand{\xmark}{\ding{55}}
\newcommand{\cmark}{\ding{51}}
\usepackage{color}
\usepackage[normalem]{ulem}
\usepackage{algorithm,algorithmicx}
\usepackage{braket}
\usepackage{bm}

\begin{document}
\ninept
\maketitle
\begin{abstract}
Despite the rapid advance of automatic speech recognition (ASR) technologies, accurate recognition of cocktail party speech characterised by the interference from overlapping speakers, background noise and room reverberation remains a highly challenging task to date. 
Motivated by the invariance of visual modality to acoustic signal corruption, audio-visual speech enhancement techniques have been developed, although predominantly targeting overlapping speech separation and recognition tasks. In this paper, an audio-visual multi-channel speech separation, dereverberation and recognition approach featuring a full incorporation of visual information into all three stages of the system is proposed. 
The advantage of the additional visual modality over using audio only is demonstrated on two neural dereverberation approaches based on DNN-WPE and spectral mapping respectively. 
The learning cost function mismatch between the separation and dereverberation models and their integration with the back-end recognition system is minimised using fine-tuning on the MSE and LF-MMI criteria. 
Experiments conducted on the LRS2 dataset suggest that the proposed audio-visual multi-channel speech separation, dereverberation and recognition system outperforms the baseline audio-visual multi-channel speech separation and recognition system containing no dereverberation module by a statistically significant word error rate (WER) reduction of 2.06 \% absolute (8.77 \% relative).
\end{abstract}

\section{Introduction}
\label{sec:intro}
Despite the rapid progress of automatic speech recognition (ASR) in the past few decades, accurate recognition of natural speech in a complex acoustic environment represented by cocktail party \cite{mcdermott2009cocktail, qian2018past} remains a highly challenging task to date. Multiple sources of interference from overlapping speakers, background noise and room reverberation lead to a large mismatch between the resulting mixed speech and clean signals. To this end, microphone arrays play a vital role in state-of-the-art ASR systems designed for overlapped and far-field scenarios \cite{barker2015third,yoshioka2015ntt, chang2019mimo}. The required acoustic beamforming techniques used to perform multi-channel array signal integration are normally implemented as time or frequency domain filters. Earlier generations of ASR systems featuring conventional multi-channel array beamforming techniques represented by either time domain delay and sum \cite{anguera2007acoustic}, or frequency domain minimum variance distortionless response (MVDR) \cite{souden2009optimal} and generalized eigenvalue (GEV) \cite{warsitz2007blind} channel integration approaches typically adopted a pipelined system architecture, containing separately developed speech separation, enhancement front-end and recognition back-end components.


With the wider application of deep learning based speech technologies, microphone array signal integration methods have evolved into a variety of DNN based designs in recent few years. These include: a) TF masking approaches \cite{bahmaninezhad2019,chen2019multi} used to predict spectral time-frequency (TF) mask labels for a reference channel that specify whether a particular TF spectrum point is dominated by the target speaker or interfering sources to facilitate speech separation; b) neural Filter\& Sum approaches directly estimating the beamforming filter parameters in either time domain \cite{sainath2017multichannel} or frequency domain \cite{xiao2016deep} to produce the separated outputs; and c) mask-based MVDR \cite{ yoshioka2018recognizing, chang2019mimo, yoshioka2018multi},  and mask-based GEV \cite{heymann2017beamnet} approaches utilizing DNN estimated TF masks to estimate target speaker and noise specific power spectral density (PSD) matrices to obtain the beamforming filter parameters, while alleviating the need of explicit direction of arrival (DOA) estimation. 

In recent years, there has been a similar trend of advancing conventional speech dereverberation approaches \cite{furuya2007,nakatani2008blind, nakatani2010speech} represented by weighted prediction error (WPE) \cite{nakatani2010speech} to their neural network based variants. These include: a) the DNN-WPE \cite{kinoshita2017neural, heymann2018frame} method, which use neural network estimated target signal PSD matrices in place of those traditionally obtained using maximum likelihood trained complex value GMMs \cite{nakatani2010speech} in the dereverberation filter estimation; and b) the complex spectral masking \cite{williamson2017time, wang2020multi} approach learning a direct TF spectral masking between reverberated and anechoic data. Furthermore, the joint optimization of neural network based speech separation, dereverberation and denoising components with a multi-channel beamforming framework \cite{kagami2018joint, boeddeker2020jointly} has been proposed as a comprehensive and overarching solution to the cocktail party speech problem, and has drawn increasing research interest. 

Motivated by the invariance of visual modality to acoustic signal corruption, and the complementary information provided on top of audio modality, there has been a long interest in developing audio-visual speech processing enhancement \cite{afouras2018conversation, wu2019time, ephrat2018looking, gu2020multi} and recognition \cite{makino2019recurrent,afouras2018deep, noda2015audio, mroueh2015deep, yu2020audio} techniques.  To date, these prior audio-visual speech processing researches were predominantly conducted in the context of either only the speech enhancement front-end \cite{afouras2018conversation, wu2019time, ephrat2018looking, gu2020multi} or the recognition back-end \cite{afouras2018deep, zhang2019robust, noda2015audio, mroueh2015deep, yu2020audio}, while a holistic, consistent incorporation of visual information in all stages of the system (speech separation, dereverberation and recognition) has not been studied. 

To address this issue, an audio-visual multi-channel speech separation, dereverberation and recognition approach featuring a full incorporation of visual information into all three stages of the system is proposed. The advantage of the additional visual modality over using audio only is demonstrated on two neural dereverberation approaches based on DNN-WPE and spectral mapping respectively. The learning cost function mismatch between the separation and dereverberation models and their integration with the back-end recognition system is minimised using fine-tuning on the MSE and LF-MMI criteria.
Experiments conducted on the LRS2 dataset suggest that the proposed audio-visual multi-channel speech separation, dereverberation and recognition system outperforms the baseline audio-visual multi-channel speech separation and recognition system without dereverberation module by a statistically significant word error rate (WER) reduction of 2.06\% absolute (8.77\% relative). 

The main contributions of this paper are summarized below: First, to the best of our knowledge, this paper presents the first use of a complete audio-visual multi-channel speech separation, dereverberation and recognition system architecture featuring a full incorporation of visual information into all three stages. In contrast, prior researches incorporate video modality in either only the speech enhancement front-end \cite{afouras2018conversation, wu2019time, gu2020multi}, recognition back-end \cite{afouras2018deep, noda2015audio, mroueh2015deep, yu2020audio}, or both multi-channel speech separation and recognition stages \cite{yu2021audio, yu2020audiomultichannel} but excluding the dereverberation component. Second, a more complete experimental validation of the advantage of audio-visual versus audio only dereverberation approaches of multiple forms (DNN-WPE, spectral mapping) is presented, as previous research \cite{tan2020audio} only considered the spectral mapping method.


\begin{figure*}[htb]
\centering
\includegraphics[width=15cm]{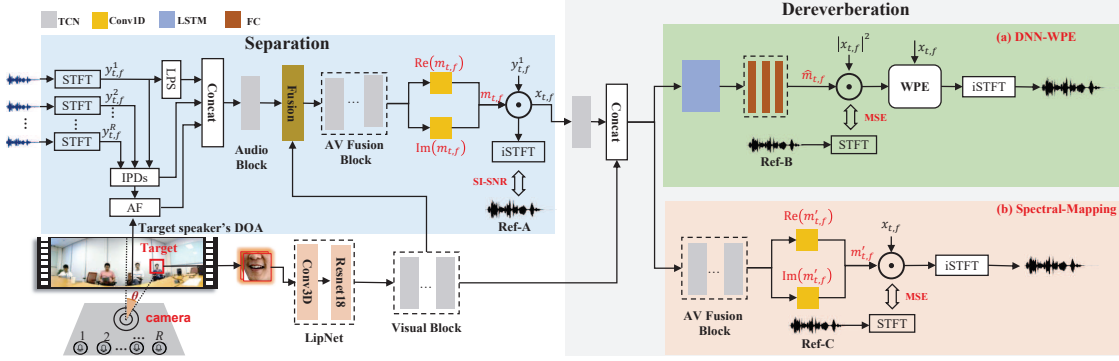}
\caption{Illustration of the audio-visual multi-channel speech separation component (top left) and DNN-WPE or spectral mapping based dereverberation modules (top right,  and bottom right respectively), where $y_{t,f}^{r}$ is the complex spectrum of each channel and $x_{t,f}$ is the separated output. $m_{t,f}$ and and $m^{\prime}_{t,f}$ are the complex ideal ratio masks of the target speaker for separation and dereverberation, where ${\rm Re}(\cdot)$ and ${\rm Im}(\cdot)$ denote the real and imaginary parts. $\hat{m}_{t,f} $ is the real-valued mask for DNN-WPE dereverberation of Eq. (\ref{eq3.5}). Ref-A, Ref-B and Ref-C denote reverberant speech, early reverberant speech and anechoic speech of the reference (1st) channel for model training respectively.}
\label{fig:pipeline}
\vspace{-1.5em}
\end{figure*}

\section{Audio-Visual Multi-channel Separation}
\vspace{-5pt}
\label{sec:sep}
This section introduces the audio-visual multi-channel separation component used before dereverberation in the overall system.  \\
\vspace{-15pt}
\subsection{Audio and visual modality inputs}
\noindent\textbf{Audio modality:}
As is illustrated in the top left corner of Figure \ref{fig:pipeline}, three types of audio features including the complex spectrum, the inter-microphone phase differences (IPDs) \cite{yoshioka2018recognizing} and location-guided angle feature (AF) \cite{chen2018multi} are adopted as the audio inputs. The complex spectrum of all the microphone array channels are first computed through short-time Fourier transform (STFT).

\noindent\textbf{IPDs features} were used to capture the relative phase difference between different microphone channels and provide additional spatial cues for TF masking based multi-channel speech separation.  These can be computed as follows:
\begin{equation}
{\rm IPD}^{(m,n)}_{t,f} = \angle({y^m_{t,f}}/{y^n_{t,f}}) 
\end{equation}
where \(y^m_{t,f}\) and \(y^n_{t,f}\) denote the STFT's TF bins of mixed speech at time $t$ and frequency bin $f$ on $m$th and $n$th microphone channels, respectively. The operator $\angle(\cdot)$ outputs the angle between them.
 
\noindent\textbf{Angle features} that are based on the approximated direction of arrival (DOA) were also incorporated to provide further spatial filtering constraint. In this work, the approximate DOA of a target speaker, $\theta$, is obtained by tracking the speaker’s face from a 180-degree wide-angle camera, as is shown in the bottom left corner of Figure \ref{fig:pipeline}. This allows the array steering vector corresponding to the target speaker to be expressed as follows:
\begin{align}
    {\bf{G}}(f) =  \left[e^{-j \phi_1\cos(\theta)}, e^{-j\phi_r\cos(\theta)}, ...,e^{-j\phi_R\cos(\theta)}\right]
\end{align}
where $\phi_r = 2\pi f d_{1r}/c$ and $d_{1r}$ is the distance between the  first (reference) and $r$th microphone ($d_{11} = 0$). $c$ is the sound velocity.
Based on the computed steering vector, the location-guided AF feature introduced in \cite{chen2018multi, gu2020multi} are also adopted to provide further discriminative information for the target speaker as follows:
\begin{align}
    \text{AF}(t,f) = \sum_{\{(m,n)\}} \frac{\big{\langle}\text{vec}\big{(}\frac{G_n(f)}{G_m(f)}\big{)}, \text{vec}\big{(}\frac{y^{m}_{t,f}}{y^{n}_{t,f}}\big{)}\big{\rangle}}{\big{\|}\text{vec}\big{(}\frac{G_n(f)}{G_m(f)}\big{)}\big{\|}\cdot \big{\|}\text{vec}\big{(}\frac{y^{m}_{t,f}}{y^{n}_{t,f}}\big{)}\big{\|}}
\end{align}
where $\|\cdot\|$ denotes the vector norm, $\langle\cdot,\cdot\rangle$ represents the inner product and $\{(m,n)\}$ denotes the selected microphone pairs.
$\text{vec}(\cdot)$ transforms the complex value into a 2-D vector, where the real and imaginary parts are regarded as the two vector components.  

Following the previous researches on audio-visual multi-channel speech separation \cite{yu2021audio, yu2020audiomultichannel}, temporal convolutional networks (TCNs) \cite{luo2019conv} are used in the speech separation module. As shown in the left corner of Figure \ref{fig:pipeline}, the log-power spectrum (LPS) features of the reference microphone channel were initially concatenated with the IPDs and AF features computed above before being fed into the TCN based audio block to compute the audio embedding. \\
\noindent\textbf{Visual modality:}
Considering the invariance of visual modality to acoustic signal corruption, and the complementary information provided on top of audio modality, the visual feature of target lip movements is extracted by a LipNet \cite{afouras2018deeplip} as shown in Figure \ref{fig:pipeline} (bottom left, in light brown). The LipNet consists a 3D convolutional layer and a 18-layer ResNet. Before fusing the visual features with the audio embedding to improve the estimation, the lip features are firstly fed into the visual block containing 5 TCNs (Figure 1, bottom left in grey) to compute the visual embedding.\\
\noindent\textbf{Audio-visual modality fusion:}
In this work, a factorised attention-based modality fusion method consistent with our previous work \cite{yu2021audio} was utilised in the separation module. This attention based fusion block (Figure 1, left middle in dark brown) combines the audio and visual embeddings from the outputs of the audio and visual TCN embedding blocks respectively. The outputs of this fusion layer are sent into the audio-visual (AV) embedding block containing 3 TCN blocks (Figure 1, left middle in grey) to compute the final audio-visual embedding features.
\vspace{-1ex}
\subsection{TF masking based speech separation}
\vspace{-0.5ex}
After modality fusion, the above audio-visual embeddings are fed into 2 Conv1D blocks (Figure 1, left in yellow) to estimate the complex ideal ratio mask (cIRM) $m_{t,f}$ of the target speech for the separation module. The estimated speech complex spectrum is:
\begin{equation}
x_{t,f} = m_{t,f} y_{t,f}^{r} \label{2.1}
\end{equation}
where $m_{t,f} \in \mathbb{C}$ is the cIRM of the target speaker, $y_{t,f}^{r}$ is the reference channel complex spectrum of mixed speech (without loss of generality, the first channel is selected as the reference channel throughout this paper), before being fed into the subsequent audio-visual dereverberation component of the following Section 3. Given the estimated complex spectrum, the time-domain separated speech can be computed by the inverse short-time Fourier transform (iSTFT) and  the SI-SNR cost function is used to optimize the separation neural networks.

\section{Audio-Visual Dereverberation}
\label{sec:dervb}

\subsection{Audio-visual DNN-WPE based dereverberation}
Let $x_{t,f}$ denotes the observed speech impaired by reverberation in the STFT domain at time frame index $t$ and frequency bin index $f$. The desired signal $\hat d_{t,f}$ can be obtained by inverse filter to subtract the estimated tail reverberation from the observed signal \cite{nakatani2010speech, kinoshita2017neural} as:
\begin{equation} \label{eq3.1}
	\begin{split}
	\hat d_{t,f} & = x_{t,f} - \sum_{\tau=D}^{D+L-1}{g^*_{\tau, f}}{x_{t-\tau, f}} = x_{t,f} - \mathbf{\hat g}_{f}^{H} \mathbf{x}_{t-D, f}
	\end{split} 
\end{equation}
where $(\cdot)^*$ and $(\cdot)^H$ operators denote the complex conjugate operator and conjugate transpose operator. $D$ denotes the prediction delay and $L$ is the filter tap. $\mathbf{x}_{t-D,f}$ and $\mathbf{\hat g}_{f}$ are vector representations of the observed signal and filter weights at frequency bin index $f$ with $L$ elements.

In conventional weighted prediction error (WPE) \cite{nakatani2010speech} dereverberation, the filter weights $\mathbf{\hat g}_{f}$ are estimated as:
\begin{align}
& \mathbf{\hat g}_{f} = {\left(\sum_{t}\frac{\mathbf{x}_{t-D, f} \mathbf{x}^H_{t-D, f} }{\lambda_{t,f}}\right)^{-1}}\left(\sum_{t}\frac{\mathbf{x}_{t-D, f} x^*_{t, f} }{\lambda_{t,f}}\right)  \label{eq3.2}
\end{align}
where $\lambda_{t,f}=| \hat d_{t,f}|^2$ is the PSD of the target speech spectrum.
In DNN-WPE, $\lambda_{t,f}$ is predicted by estimating the TF mask as follows:
\begin{equation}
 \lambda_{t,f} = \hat{m}_{t,f} \left | x_{t,f} \right|^2 \label{eq3.5}
\end{equation}
where $\hat{m}_{t,f} \in \mathbb{R} $ is the estimated TF mask produced by the DNN-WPE module (Figure 1, top right in green) using audio embeddings alone, or optionally fused audio-visual features as the input to leverage the invariance of video modality to reverberation in this work.

During training, the spectrum of the early reverberant speech (with first 50ms reverberation following \cite{heymann2018frame}) is adopted as the training target. The dereverberation module is optimized using the mean square error (MSE) cost. 
Once the mask is predicted, $\mathbf{\hat g}_{f}$ can be estimated to facilitate dereverberation.
\vspace{-1ex}
\subsection{Audio-visual spectral mapping for dereverberation}
\vspace{-0.5ex}
In addition to the audio-visual DNN-WPE approach, an audio-visual spectral mapping approach is also proposed. As shown in Figure \ref{fig:pipeline} (bottom right corner, in pink), the LPS of separated speech is firstly fed into the audio block to compute the audio embeddings and then concatenated with the lip embedding extracted from the visual block. The concatenated audio-visual embedding are then forwarded into the AV fusion block. Finally, the audio-visual embeddings are forwarded into 2 Conv1D blocks to estimate the cIRM $m^{\prime}_{t,f}$ of the non-reverberant target speech. The estimated speech complex spectrum can be obtained using Eq.(\ref{2.1}). During training, the spectrum of the anechoic speech is used as the training target and MSE is adopted as the loss function. 

\vspace{-5pt}
\section{Integration of enhancement front-end \& recognition back-end}
\vspace{-5pt}
When designing cocktail part speech recognition systems, the separation, dereverberation and recognition components are often used in a pipelined manner. However, the performance of such system can be sub-optimal due to the mismatch among the different training error costs used by the three components. To this end, a tighter integration between system components, for example, the separation and dereverberation modules, can be achieved via joint fine-tuning on the dereverberation MSE cost alone, or an interpolated SI-SNR and MSE error loss function. 
\footnote{In this paper, we only show the MSE loss joint fine-tuning results since using the multi-task loss gives limited performance improvement.} Their further integration of the audio only or audio-visual CLDNN based back-end recognition component was performed by fine-tuning using the LF-MMI sequence training criterion \cite{yu2021audio} given the enhanced outputs. 

\begin{table}[h]
\centering
\vspace{-20pt}
\caption{Performance of single channel ASR and AVSR systems trained and evaluated on anechoic, early-reverberant, reverberant and reverberant-noisy-overlapped speech. ${\dag}$ denotes a statistically significant difference obtained over the  baseline (sys. 1, 3, 5, 7).}
\label{table:1}
\scalebox{0.8}{
\begin{tabular}{c|c|c|c} 
\toprule[1pt]
Sys & Data & +visual & WER(\%) \\
\hline
1 & \multirow{2}{*}{anechoic} & \xmark & 13.87  \\ 
2& &\cmark & 12.28$^{\dag}$   \\
\hline
3 & \multirow{2}{*}{early-reverberant} & \xmark & 18.29 \\ 
4 & & \cmark & 13.51$^{\dag}$  \\
\hline
5 & \multirow{2}{*}{reverberant} & \xmark & 23.35 \\ 
6 & & \cmark & 17.61$^{\dag}$  \\
\hline
7 & \multirow{2}{*}{overlapped-noisy-reverberant} & \xmark & 84.01\\ 
8 & & \cmark & 42.25$^{\dag}$  \\
\hline
\end{tabular}}
\vspace{-10pt}
\end{table}

\begin{table*}[h!]
\centering
\vspace{-20pt}
\caption{Separation, dereverberation and recognition on LRS2 overlapped-noisy-reverberant dataset. `Sep.', `Dervb.', `Recg.' denotes separation, dereverberation and recognition. `Anec.' denotes ASR system trained on anechoic speech. `AF' and `SpecM' denotes angle feature and spectral mapping. FT denotes fine tuning. ${\star}$, ${\dag}$  and ${\ddag}$ denotes statistically significant difference over baseline (sys. 1,2,3). }
\label{table:3}
\resizebox{150mm}{29mm}{
\begin{tabular}{c|c|c|c|c|c|c|c|c|c|c|c|c|c} 
 \toprule[1pt]
 \multirow{2}{*}{Sys} & \multicolumn{2}{c|}{Sep.} & \multicolumn{2}{c|}{Dervb.} & Recg. & Anec. & \multicolumn{3}{c}{Pipeline} &  \multicolumn{3}{|c|}{Jointly FT (Sep.+ Dervb.)} & FT back-end \\
 \cline{2-14} 
 & \footnotesize AF & +visual &  \footnotesize method  &  +visual  & +visual & \footnotesize WER & \footnotesize PESQ & \footnotesize SRMR & \footnotesize WER  & \footnotesize PESQ & \footnotesize SRMR &  \footnotesize WER & \footnotesize WER\\
 \hline\hline
 \multicolumn{6}{c|}{raw 1 channel reverberant-noisy-overlapped} & 85.71  & 1.65 & 4.01& 84.01 & \multicolumn{3}{c|}{-} & - \\
 \hline
 1 & \cmark & \xmark  & \multicolumn{2}{c|}{-}  & \xmark &  53.29 & 2.30 & 6.20 & 39.38 & \multicolumn{3}{c|}{-} & - \\
 \hline
 2 & \cmark & \cmark  & \multicolumn{2}{c|}{-} & \xmark & 48.51 & 2.37 & 6.44 & 36.46 & \multicolumn{3}{c|}{-} &  -  \\
 \hline
 3 & \cmark & \cmark  & \multicolumn{2}{c|}{-} & \cmark & \bf{33.23} &  2.37 & 6.44 &  \bf{24.92} & \multicolumn{3}{c|}{-}  & \bf{23.50}  \\
 \hline
 \hline
 4  & \cmark & \xmark  & \multirow{2}{*}{ \footnotesize DNN-WPE}  & \xmark & \xmark &51.76$^\star$& 2.32 & 6.65 & 38.38 & 2.32 & 6.70 & 38.21$^\star$ & - \\

 5  & \cmark & \xmark  &  & \cmark & \xmark & 51.25$^\star$ & 2.33 & 6.70  &  37.96$^\star$& 2.33 & 6.78 & 37.24$^\star$ & -  \\
  \hline
 6  & \cmark & \xmark & \multirow{2}{*}{\footnotesize SpecM}   & \xmark & \xmark & 51.94 & 2.37 & 7.79 & 38.08$^\star$ & 2.39 & 8.33 &  37.63$^\star$ & - \\

 7  & \cmark & \xmark  & & \cmark & \xmark & 51.68$^\star$ & 2.38 & 8.00 & 36.98$^\star$ & 2.39 & 8.45 &  36.58$^\star$ & -\\
 \hline
 \hline
 8 & \cmark & \cmark & \multirow{2}{*}{\footnotesize DNN-WPE}  & \xmark & \xmark  & 47.60 & 2.39  & 6.78 & 35.09$^\dag$ & 2.40 & 6.90 & 34.52$^\dag$ & -   \\
 9 & \cmark & \cmark  &  &\cmark & \xmark & 47.55$^\dag$ & 2.40 & 6.73 & 34.58$^\dag$ & 2.41 & 6.93 & 33.69$^\dag$ & - \\
 \hline
 10 & \cmark & \cmark  & \multirow{2}{*}{\footnotesize SpecM}   &\xmark & \xmark & 47.45$^\dag$ & 2.46 & 7.55 & 34.31$^\dag$ & 2.47 & 8.68 & 34.05$^\dag$ & -\\
 11 & \cmark & \cmark  &  &\cmark & \xmark & \bf{46.88}$^\dag$ & 2.48 & 7.77 & \bf{33.44}$^\dag$ & 2.49 & 8.71 & \bf{32.91}$^\dag$ & - \\
 \hline
 \hline
 12 & \cmark & \cmark & \footnotesize DNN-WPE  &\cmark & \cmark & 32.15$^\ddag$ & 2.40 & 6.73 & 23.69$^\ddag$ & 2.41 & 6.93 & 22.99$^\ddag$ & 21.91$^\ddag$  \\
 \hline
 13 & \cmark & \cmark & \footnotesize SpecM  &\cmark & \cmark & \bf{31.32}$^\ddag$  &  2.48 & 7.77 &  \bf{22.72}$^\ddag$ &  2.49 & 8.71  &  \bf{22.38}$^\ddag$ & \bf{21.44}$^\ddag$ \\
 \hline
\end{tabular}
\vspace{-2em}
}
\end{table*}

\section{Experiment \& Results}
\vspace{-5pt}
\subsection{Experiment Setup} 
\textbf{Simulated mixed speech}: The multi-channel overlapped-noisy-reverberant speech is simulated using the LRS2 dataset. A 15-channel symmetric linear array described in \cite{yu2021audio} is used in the simulation process. 843-point source noises and 40000 Room Impulse Responses (RIRs) generated by the image method \cite{habets2006room} in 400 different simulated rooms were used in our experiment. The distance between a sound source and the microphone array center is randomly sampled from the range of 1m to 5m and the room size is ranging from 1m $\times$ 1m $\times$ 2m to 30m $\times$ 30m $\times$ 5m (length $\times$ width $\times$ height). The reverberation time $T_{60}$ is sampled from a range of 0.06s to 1.12s. The average overlapping ration is around 85\%. The SNR is randomly chosen from 0, 5, 10, 15 and 20 dB, and SIR from -6, 0 and 6 dB. The simulated dataset is split into three subsets with 91.7k, 2.2k, and 1.2k utterances respectively for training, validation and test. Statistical significance test was conducted at level $\alpha = 0.05$ based on matched pairs sentence segment word error (MAPSSWE) for recognition performance analysis


\noindent{\textbf{Implementation details:}}
Details of the IPD, AF features and the baseline audio-visual back-end ASR can be found in \cite{yu2021audio}.
For DNN-WPE dereverberation, the prediction delay D and filter tap L were set to 3 and 18, respectively. The number of iterations for parameter estimation in conventional WPE was set to 3. \\

\begin{table}[h!]
\centering
\vspace{-20pt}
\caption{Performance of ASR systems trained on anechoic and dereverberated speech. `Anec.' denotes ASR systems trained on anechoic speech. `Retrain.' denotes the ASR systems trained on the dereverberated data. ${\dag}$ denotes a statistically significant difference obtained over the corresponding audio-only baselines (sys. 2, 4).}
\label{table:2}
\scalebox{0.8}{
\begin{tabular}{c|c|c|c|c|c|c} 
 \toprule[1pt]
 \multirow{2}{*}{Sys} & \multicolumn{2}{c|}{Dervb.} & \multirow{2}{*}{\scriptsize PESQ} & \multirow{2}{*}{\scriptsize SRMR} &  \multicolumn{2}{c}{WER(\%)} \\
\cline{2-3} 
\cline{6-7}
 &  \footnotesize method  &  +visual &  &  & \footnotesize Anec. & \scriptsize Retrain. \\
 \hline\hline
 \multicolumn{3}{c|}{- raw reverberant -} & 2.76  & 5.85 & 34.29 & 23.35 \\
 \hline
 1 & \scriptsize  WPE \cite{nakatani2010speech} & - & 2.77  &  5.90 & 32.34 & 21.70 \\
 \hline
 2 &  \multirow{2}{*}{ \scriptsize DNN-WPE} & \xmark & 2.79 & 6.54 &  28.03 & 19.56 \\
 3 &  & \cmark & 2.83 &  6.58 & 27.64  &  18.87 $^{\dag}$ \\
 \hline
 4 & \multirow{2}{*}{\scriptsize SpecM } & \xmark & 2.92  & 7.67 & 27.94 & 19.52 \\
 5 &  &  \cmark & 2.97 &  8.05 & 27.53$^{\dag}$ & 18.53$^{\dag}$  \\
 \hline
\end{tabular}
\vspace{-10pt}
}
\end{table}

\vspace{-2.5em}
\subsection{Experiment Results}
\noindent\textbf{Speech recognition without front-end: }
Table \ref{table:1} presents the WERs of various LF-MMI based ASR and AVSR systems that contains no separation and dereverberation components, trained and evaluated on four types of data: anechoic, early-reverberant, reverberant and overlapped-noisy-reverberant speech. It can be observed that using visual information can significantly improve the recognition performance over the audio-only systems. In particular, the audio-visual recognition system trained on overlapped-noisy-reverberant speech outperforms the audio-only system by up to 41.76 \% (sys.7 vs sys.8) absolute WER reduction on test data of the same condition. \\
\noindent\textbf{Performance of audio-visual dereverberation:}
The performance of various dereverberation approaches on reverberant only (non-overlapping) speech are evaluated on ASR systems constructed using anechoic or the corresponding dereverberated speech, and shown in Table \ref{table:2}. It can be observed that adding visual information in DNN-WPE and spectral mapping (SpecM) dereverberation produced consistent improvements in terms of PESQ \cite{recommendation2001perceptual}, speech to reverberation modulation energy ratio (SRMR) \cite{falk2010non} and WER (sys. 2 vs sys. 3,  sys. 4 vs sys. 5). \\

\vspace{-0.2cm}
\noindent\textbf{Performance on overlapped-noisy-reverberant speech:}
Their performance are further evaluated on audio-visual multi-channel speech separation, dereverberation system constructed using the overlapped-noisy-reverberant data in Table \ref{table:3}, with either a partial, or full incorporation of visual information into all three stages. Several trends can be observed from Table \ref{table:3}: 
1) Using visual feature in both the separation and recognition components significantly improves the recognition performance on both the anechoic speech trained `Anec.' and pipeline systems by up to 20.06 \% and 14.46 \% (sys. 1 vs sys. 3), as well as the PESQ and SRMR scores. 
2) In contrast to the pipeline systems only containing separation front-ends (sys. 1, 2, 3), adding audio-only or audio-visual dereverberation modules (DNN-WPE \& SpecM) produced WER reductions ranging from 1\% to 2.4\% (sys. 4, 5, 6, 7 vs sys. 1), 1.37\% to 3.02 \% (sys. 8, 9, 10, 11 vs sys. 2 ), 1.23\% to 2.2\% (sys. 12, 13 vs sys. 3). Among these, leveraging visual modality in dereverberation module consistently produced better performance with WER reductions up to 1.1\% absolute (sys.7 vs sys. 6).
3) Further jointly fine-tuning the separation and dereverberation components using dereverberation MSE cost function (col. ``Jointly FT (Sep.+ Dervb.)", Table \ref{table:3}) produced PESQ and SRMR improvements of 0.04-0.12 and 0.53-2.27 points (sys. 12, 13 vs sys.3), in addition to consistent WER reductions over the pipeline systems. 
4) Finally, to tightly integrate the enhancement front-end and recognition back-end, fine-tuning three AVSR systems (sys. 3, 12, 13) on their respective enhanced speech outputs (last col. Table \ref{table:3}) further reduced the WER while retaining the same trend. The final fine-tuned AVSR systems with an audio-visual dereverberation module outperform the baseline AVSR system  by 1.59\%-2.06\% in WER reduction (sys. 12, 13 vs sys.3).

\vspace{-8pt}
\section{Conclusions}
\vspace{-8pt}
In this paper, an audio-visual multi-channel speech separation, dereverberation and recognition approach featuring a full incorporation of visual information into all three stages of the system is proposed. The advantages of visual modality over using acoustic features only is demonstrated on two neural dereverberation approaches based on DNN-WPE and spectral mapping using the LRS2 dataset simulated speech containing overlapping, noise and reverberation. Future research will focus on improving the integration between the separation, dereverberation and recognition components.    

\label{sec:Conclusions} 
\vspace{-8pt}
\section{ACKNOWLEDGEMENTS}
\vspace{-8pt}
This research is supported by Hong Kong RGC GRF grant No. 14200218, 14200220, 14200021, TRS T45-407/19N, and Innovation \& Technology Fund grant No. ITS/254/19.

\vfill\pagebreak

\bibliographystyle{IEEEbib}
\footnotesize
\bibliography{strings,refs}

\end{document}